\documentclass{elsart}

\journal{Chemical Physics}

\usepackage{epsfig}

\begin{document}

\begin{frontmatter}



\title{Strange kinetics: conflict between density and trajectory description}

\author{M. Bologna}
\ead{mb0015@unt.edu }

\address{ Center for Nonlinear Science, University of North\\
Texas, P.O. Box 311427, Denton, Texas 76203-5370, USA\\
}
\author{ P. Grigolini \corauthref{cor}}

\address{ Center for Nonlinear Science, University of North\\
Texas, P.O. Box 311427, Denton, Texas 76203-5370, USA\\
}
\address{ Dipartimento di Fisica dell$^{\prime }$Universit\`{a} di Pisa \\
and INFM, Piazza Torricelli 2, 56127 Pisa, Italy\\
}
\address{  Istituto di Biofisica CNR, Area della Ricerca di \\
Pisa, Via Alfieri 1, San Cataldo 56010 Ghezzano-Pisa, Italy\\}

\ead{grigo@unt.edu }

\author{ B. J. West}
\address{US Army Research Office, Research Triangle Park,\\
NC 27709, USA}

\ead{westb@aro-emh1.army.mil }

\corauth[cor]{Corresponding author.}

\begin{abstract}
We study a process of anomalous diffusion, based on intermittent velocity
fluctuations, and we show that its scaling depends on whether we observe the
motion of many independent trajectories or that of a Liouville-like equation
driven density. The reason for this discrepancy seems to be that the
Liouville-like equation is unable to reproduce the multi-scaling properties
emerging from trajectory dynamics. We argue that this conflict between
density and trajectory might help us to define the uncertain border between
dynamics and thermodynamics, and that between quantum and classical physics
as well.
\end{abstract}


\end{frontmatter}




\date{\today}     
\maketitle

\date{}

\section{Introduction}

In statistical physics we assume two distinct, but equivalent descriptions
of the dynamics of physical systems; the trajectories of individual
particles and the density function for an ensemble of such trajectories. The
dynamics of the trajectories are determined by Hamilton's equations of
motion and the evolution of the phase space density is determined by
Liouville's equation. The relation between the two pictures is similar to
that between the Heisenberg and Schr\"{o}dinger representations. The word
representation clearly indicates that these are just two ways of looking at
the same physical phenomenon and it would therefore appear frivolous to
question their equivalence. In this paper we do call this equivalence into
question. However, we do not elect to formulate a grand theory to establish
the difference between the two points of view, this has been done by
Petrosky and Prigogine \cite{petrosky,petrosky2}, but rather we take the
more modest approach of establishing an inconsistency based on a simple
physical problem.

First let us examine how the connection between trajectories, $x\left(
t\right) ,$ and phase space densities, $\sigma \left( x,t\right) $, are
usually made. A numerical treatment of the phase space density requires, by
necessity, a transformation of the Liouville density function $\sigma \left(
x,t\right) $ into a probability, through the relation

\begin{equation}
p\left( x_{i},t\right) =\left\langle \sigma \left( x_{i},t\right)
\right\rangle \delta x_{i}  \label{prob}
\end{equation}
where the $x$-axis is divided into $N$ equal parts indexed by $j=1,2,...,N$, 
$\delta x_{i}$ is a small phase space interval $\left( x_{i},x_{i}+\delta
x_{i}\right) $ and the probability density is normalized to unity at all
times

\begin{equation}
\sum_{i=1}^{N}p\left( x_{i},t\right) =1.  \label{normalization}
\end{equation}
In statistical mechanics $p\left( x_{i},t\right) $ is the probability of the
system being in the interval $\left( x_{i},x_{i}+\delta x_{i}\right) $ at
time $t.$

As a practical matter, in numerical simulations, the quantity $p\left(
x_{i},t\right) $ is the histogram defined by counting the number of
trajectories that lie in the interval $\left( x_{i},x_{i}+\delta
x_{i}\right) $ at time $t$. It is clear that this interval can be made as
small as we wish, as long as it remains non-zero in numerical calculations.
If we consider a set of initial conditions that produce a sharply peaked
Liouville density, the density could either remain peaked as it wanders
through phase space, or it could spread out, thereby producing the
continuous spectrum of $\sigma \left( x,t\right) $. There are at least two
ways such spreading can occur. The first is by the action of infinitely many
degrees of freedom in the system. The second is by trajectory instabilities,
that being, chaos produced by nonlinear interactions among the degrees of
freedom. The latter situation gives rise to dynamically generated random
trajectories, whereas the former does not. The kind of statistics emerging from the former case
 is the result of an assumption, one that is made
regarding the initial states of the system. Thus, statistics in the latter
case is a consequence of dynamics whereas in the former case it is the
result of an assumption.

It is important to stress that the theoretical concept of probability, or
Liouville density, and computational probability are not equivalent. In fact
while the theoretical probability could be a very complex function, the
computational probability is essentially a relative frequency concept,
strongly dependent on the size of the cells that are used in the
calculation. The computational probability can be interpreted more as a
coarse-grained probability than as a true Liouville density. Consequently,
it is not unexpected that the two pictures, that of trajectories and that of
densities, at least in principle, might produce different results.

However, we know that trajectories and densities have historically produced
the same statistical effects. We know, for example, that when the Markov
condition applies, the probabilistic representation, based on trajectories,
follows the Central Limit Theorem (CLT). This agrees with the density perspective. In fact, as recently as last year,
Lee \cite{lee} showed that the adoption of the Liouville approach is
compatible, in the long-time limit, with Fick's law, and, consequently, with
the tenets of the CLT. In this paper we show that this equivalence is lost
when the Generalized CLT (GCLT) of L\'{e}vy-Gnedenko \cite{kolmogorov}
applies. In this latter case, the density picture cannot properly reproduce
the mechanism of memory erasure necessary for the  GCLT to apply. The
density representation seems to be more adequate to mimic the relaxation
processes, determined by the action of many degrees of freedom, without
involving trajectory randomness, and therefore compatible with the existence
of an infinite memory.

\section{On a time convoluted equation with infinite memory}

In this section we derive a single equation of motion for the Liouville
density, in two distinct ways. The first is based on the Liouville equation
of all the variables, both the diffusing variable $x$ and those responsible
for the fluctuations generating the diffusion process under study. The
second way is  based on the direct integration of the equation of motion
for a single trajectory, and serves essentially the purpose of double
checking the generalized diffusion equation resulting from the first method.
The first method, identical to that used in an earlier publication\cite
{allegrini}, is applied imagining that fluctuations are produced by a set of
bath variables, and is compatible with the theoretical perspective, on the
foundation of statistical mechanics, based on the action of infinitely many
degrees of freedom. However, as shown in Appendix, the same result can be
obtained by means of trajectory randomness.

\subsection{\protect\bigskip Liouville argument}

We consider one of the simplest differential equations

\begin{equation}
\frac{dx\left( t\right)}{dt}=\xi \left( t\right) ,  \label{lou1}
\end{equation}
where $\xi \left( t\right) $ is a two state random process taking the values 
$\pm W$. If $\phi \left( x,\xi ,t\right) $ is the phase space distribution
function, then the equation of evolution \ corresponding to the dynamical
equation (\ref{lou1}) is

\begin{equation}
\frac{\partial }{\partial t} \phi \left( x,\xi ,{\bf R}, t\right) =\left( -%
\widehat{\xi }\frac{\partial }{\partial x}+\widehat{\Gamma }\right) \phi
\left( x,\xi,{\bf R},t\right) .  \label{lou2}
\end{equation}
We are adopting a quantum-like formalism. Thus, $\widehat{\Gamma }$ is an
operator characterizing the dynamics of the $\xi -$process and $\widehat{\xi 
}$ is an operator having the eigenvalues $\pm W$, namely 
\begin{equation}  \label{plusandminus}
\hat \xi|\pm> = \pm W|\pm>.
\end{equation}
The underlying process generating $\xi \left( t\right) $ need not be
specified, but one realization of it could be a Hamiltonian system with a
set of variables {\bf R}. These latter variables can be infinitely many so
as to result in the relaxation of the correlation properties of the system.

At equilibrium, the two states $|+>$ and $|->$ must have the same
statistical weight. Thus we assume that the bath equilibrium corresponds to
the state 
\begin{equation}  \label{equilibriumofxi}
|p_{0}> = \frac{1}{\sqrt{2}}(|+>+|->) \Pi({\bf R}),
\end{equation}
where $\Pi({\bf R})$ denotes the equilibrium distribution of the  variables
responsible for the stochastic dynamics of the variable $\xi$. The state $%
|p_{0}>$ is one of the eigenstates  of the operator $\hat\Gamma$. In fact, we
set 
\begin{equation}
\hat\Gamma |\mu> = -\Lambda_{\mu}|\mu>,  \label{batheigenstates}
\end{equation}
and $|p_{0}> = |\mu = 0>, \Lambda_{0} = 0$.  Within this quantum-like
formalism the variable $\xi$, as  earlier said, corresponds to the operator $%
\hat{\xi}$. This  operator, applied to the equilibrium state $|p_{0}>$
yields  the excited state 
\begin{equation}  \label{excitedstate}
|p_{1}> = \frac{(\hat{\xi}|p_{0}>)\Pi({\bf R})}{W}.
\end{equation}
This means that the operator $\hat{\xi}$ does not affect the  distribution
of ${\bf R}$. It has the effect of making the  transitions $|+>+|->
\rightarrow |+>-|->$ and $|+>-|->  \rightarrow |+>+|->$, without affecting
the other bath variables.  The ``excited'' state $|p_{1}>$ is not an
eigenstate of $\hat\Gamma$, but it is a linear combinations of the states $%
|\mu>$, with $\mu \neq 0$.  The operator $\hat{\Gamma}$ applied to the
``excited'' state $|p_{1}>$ has the effect of realizing a relaxation by
coupling the  state $|p_{1}>$ to infinitely many other eigenstates $|\mu>$. 
The correlation function $<\xi \xi(t)>$, within this quantum-like 
formalism, reads 
\begin{equation}
<\xi \xi(t)> = <p_{0}|\hat \xi exp(\Gamma t) \hat \xi |p_{0}>.  \label{xixi}
\end{equation}
On the basis of the properties of the operators $\hat \xi$ and $\hat \Gamma$%
, this correlation function can also be expressed  under the form: 
\begin{equation}
<\xi \xi(t)> = W^{2} \sum_{\mu \neq 0} <p_{1}|\mu><\mu|p_{1}>exp{%
(-\Lambda_{\mu}t)}.  \label{xixi2}
\end{equation}
It is  convenient to define 
\begin{equation}
\sigma_{\mu}(x,t) \equiv <\mu|\phi(x,\xi,{\bf R},t)> ,  \label{definitions}
\end{equation}
with $\mu =0,1,2,\ldots$. We are interpreting the  distribution $\phi$ of
Eq.(\ref{lou2}) as a sort of ket  vector $|\phi>$. By multiplying Eq. (\ref
{lou2}) on the left by  the states $|\mu>$ we get 
\begin{equation}  \label{zero}
\frac{\partial}{\partial t} \sigma_{0}(t) = -W \sum_{\mu \neq 0}^{\infty}
a_{\mu} \frac{\partial}{\partial x} \sigma_{\mu}(x,t)
\end{equation}
and, for $\mu > 0$, 
\begin{equation}  \label{mu}
\frac{\partial}{\partial t} \sigma_{\mu}(t) = -W \sum_{\mu \neq 0}^{\infty}
a_{\mu}^{*} \frac{\partial}{\partial x} \sigma_{0}(x,t) - \Lambda_{\mu}
\sigma_{\mu}(x,t),
\end{equation}
with $a_{\mu}= <\mu|\hat \xi |p_{0}>$.

Let us make the assumption that at $t = 0$ all the $\sigma_{\mu}$'s  but the
one with $\mu = 0$ vanish. This condition is equivalent  (see also Appendix)
to assuming the spatial distribution to be  statistically independent of the
``velocity'' distribution, and  has the nice effect, as we shall see below,
of resulting in an  equation of motion without an inhomogeneous term.  By
solving Eq.(\ref{mu}),  and replacing the solution into Eq. (\ref{zero}). we
get 
\begin{equation}  \label{almost}
\frac{\partial}{\partial t} \sigma_{0}(t) = W^{2} \sum_{\mu \neq 0}^{\infty}
a_{\mu} \frac{\partial}{\partial x} |a_{\mu}|^{2}
\int_{0}^{t}dt^{\prime}exp[-\Lambda_{\mu}(t-t^{\prime})] \frac{\partial^{2}}{%
\partial x^{2}} \sigma_{0}(x,t^{\prime}).
\end{equation}
From now on we shall focus on the reduced density matrix $\sigma_{0}(x,t)$
and for the sake of simplicity we shall omit the  subscript $0$. Using Eq.(%
\ref{xixi2}) we can rewrite Eq.  (\ref{almost}) in the form

\begin{equation}
\frac{\partial}{\partial t} \sigma \left( x,t\right)=\int_{0}^{t}\left
\langle \xi \left( t\right) \xi \left( t^{\prime }\right) \right\rangle 
\frac{\partial ^{2}\sigma \left( x,t^{\prime }\right) }{\partial x^{2}}
dt^{\prime }.  \label{lou6}
\end{equation}

In the case where the correlation function in (\ref{lou6}) is an exponential

\begin{equation}
\left\langle \xi \left( t\right) \xi \left( t^{\prime }\right) \right\rangle
=\left\langle \xi ^{2}\right\rangle e^{-\gamma \left( t-t^{\prime }\right) },
\label{lou7}
\end{equation}
taking the time derivative of (\ref{lou6}) yields

\begin{equation}
\frac{\partial ^{2}\sigma \left( x,t\right) }{\partial t^{2}}+\gamma \frac{%
\partial \sigma \left( x,t\right) }{\partial t}-\left\langle \xi
^{2}\right\rangle \frac{\partial ^{2}\sigma\left( x,t\right) }{\partial x^{2}%
}=0.  \label{lou8}
\end{equation}
This is the celebrated telegrapher's equation, whose phenomenological
pedigree originates with Maxwell. His (Maxwell's) argument was to include
relaxation into the wave equation and did not require the invocation of
microscopic dynamics. However, this use of dissipation was compatible with
the action of infinitely many degrees of freedom in the medium supporting
the wave motion, and made the Poincar\'{e} recurrence times of the dynamics,
infinitely long. On the other hand, the case of ordinary statistical
mechanics is where we do not find any conflict between the adoption of a
density picture and a trajectory picture.

\subsection{Trajectory argument}

The trajectory for an unforced system rate equation driven by the random
function $\xi \left( t\right) ,$ such as (\ref{lou1}), is given by

\begin{equation}
x\left( t\right) = x\left( 0\right) +\int_{0}^{t}\xi \left( t^{\prime
}\right) dt^{\prime }.  \label{traj1}
\end{equation}
We again assume that the random driving function can only take on two
values, $\pm W$. The fluctuation $\xi(t)$ gets one of these two values for
an unpredictable amount of time and in Section II C we shall see that the
the condition of anomalous diffusion here under study is due to the inverse
power nature of the corresponding distribution of sojourn times in each of
these two states. In the absence of bias, the odd-order moments of the
fluctuations vanish, and using the dichotomous nature of $\xi \left(
t\right) $ we obtain the factorization result

\begin{equation}
\left\langle \xi \left( t_{1}\right) \xi \left( t_{2}\right) \cdot \cdot
\cdot \xi \left( t_{2n-1}\right) \xi \left( t_{2n}\right) \right\rangle
=\left\langle \xi \left( t_{1}\right) \xi \left( t_{2}\right) \right\rangle
\cdot \cdot \cdot \cdot \left\langle \xi \left( t_{2n-1}\right) \xi \left(
t_{2n}\right) \right\rangle .  \label{traj2}
\end{equation}
With these conditions on the correlation properties of the random force,
introducing the correlation function

\begin{equation}
\Phi _{\xi }\left( t_{1}-t_{2}\right) =\frac{\left\langle \xi \left(
t_{1}\right) \xi \left( t_{2}\right) \right\rangle }{\left\langle \xi
^{2}\right\rangle },  \label{traj3}
\end{equation}
and using $\left\langle \xi ^{2}\right\rangle =W^{2}$, it is straight
forward to show that the moments of the system response has the convolution
form

\begin{eqnarray}
\left\langle x\left( t\right) ^{2n}\right\rangle &=&\left( 2n\right)
!W^{2n}\int_{0}^{t}dt_{1}\int_{0}^{t_{1}}dt_{2}\cdot \cdot \cdot
\int_{0}^{t_{2n}}dt_{2n-1} \times \Phi _{\xi }\left( t_{1}-t_{2}\right)
\cdot \cdot \cdot \Phi _{\xi }\left( t_{2n-1}-t_{2n}\right) .  \label{traj4}
\end{eqnarray}

Let us consider the Liouville-like equation

\begin{equation}
\frac{\partial \sigma \left( x,t\right) }{\partial t}=W^{2}\int_{0}^{t}\Phi
_{\xi }\left( t-t^{\prime }\right) \frac{\partial ^{2}\sigma \left(
x,t^{\prime }\right) }{\partial x^{2}}dt^{\prime }.  \label{traj5}
\end{equation}
This is the same equation as that derived from the quantum-like formalism of
Section IIA, Eq. (\ref{lou6}). Let us multiply it on the left by $x^{2n}$
and let us integrate over all phase space yields. After two integrations by
parts, we get

\begin{equation}
\frac{d\left\langle x^{2n};t\right\rangle }{dt}=W^{2}n(n-1)\int_{0}^{t}\Phi
_{\xi }\left( t-t^{\prime }\right) \left\langle x^{2n-2};t^{\prime
}\right\rangle dt^{\prime }.  \label{traj6}
\end{equation}
Thus, we can see that following an integration over time we obtain the
recursive moment relation 
\begin{equation}
\left\langle x^{2n};t\right\rangle
=W^{2}n(n-1)\int_{0}^{t}dt_{1}\int_{0}^{t_{1}}\Phi _{\xi }\left(
t_{1}-t_{2}\right) \left\langle x^{2n-2};t_{2}\right\rangle dt_{2},
\label{traj7}
\end{equation}
whose solution, obtained by reinsertion of the equation back into itself, is
given by (\ref{traj4}). Thus, we have the equivalence between the moments of
the dynamical variable and the moments of the phase space variable:

\begin{equation}
\left\langle x\left( t\right) ^{2n}\right\rangle =\left\langle
x^{2n};t\right\rangle .
\end{equation}
The meaning of this result is that Eq. (\ref{lou6}) is the proper
Liouville-like equation for the dynamic process under study. So, the
conflict between trajectory and density, which is the central issue of this
paper, cannot be ascribed to the fact that the density equation of motion is
questionable. It is not so, the generalized diffusion equation of Eq. (\ref
{lou6}) generates the same moments as the trajectories. The discrepancy
between the two pictures have a deep motivation that we can already try to
identify at this level. Paraphrasing Fox\cite{ronald}, we note that the
Liouville-like equation of Eq. (\ref{lou6}) is not a {\em bona fide}
diffusion equation. This is so because of its irretrievably Non-Markov
character that forces us to adopt it always with the same initial condition,
the distribution of the coordinate $x$ independent of that of "velocity" $\xi
$. As we shall see, at long times the adoption of the trajectory perspective
has, on the contrary, the effect of erasing any form of memory. This is the
true reason of the conflict between the two perspectives. The ambiguity
resulting from the constraints posed by the moments cannot be invoked. In
fact, it is straightforward to show that the moment hierarchy of Eq. (\ref
{traj7}) fits the Hamburger condition\cite{simon} ensuring the uniqueness of
the solution of Eq. (\ref{lou6}).

\subsection{Phase-space equation of motion}

The equation of motion for the Liouville density that we obtained using the
quantum-like Liouville approach, in a complete accordance with the moment
constraints resulting from the trajectory argument, is of the form

\begin{equation}
\frac{\partial \sigma \left( x,t\right) }{\partial t}=\int_{0}^{t}\Phi
\left( t-t^{\prime }\right) \frac{\partial ^{2}\sigma \left( x,t^{\prime
}\right) }{\partial x^{2}}dt^{\prime } = \int_{0}^{t}\Phi \left(t^{\prime
}\right) \frac{\partial ^{2}\sigma \left( x,t -t^{\prime }\right) }{\partial
x^{2}}dt^{\prime }.  \label{phase1}
\end{equation}
In the case when the  correlation function $\Phi \left( t\right) $ is
integrable, using the last term of the equality of (\ref{phase1}), we can
easily make the Markov approximation. This is based on replacing $\frac{%
\partial ^{2}\sigma \left( x,t -t^{\prime }\right) }{\partial x^{2}}$ with $%
\frac{\partial ^{2}\sigma \left( x,t \right) }{\partial x^{2}}$ and in
extending the time integration, from $0$ to $\infty$ rather than from $0$ to 
$t$. Thus we get

\begin{equation} 
\frac{\partial \sigma \left( x,t\right) }{\partial t}=D\frac{\partial
^{2}\sigma \left( x,t\right) }{\partial x^{2}} ,  \label{phase2}
\end{equation}
where the diffusion coefficient $D$ is given by

\begin{equation}
D=W^{2} \tau_{C}  \label{phase3}
\end{equation}
and 
\begin{equation}
\tau _{C}=\int_{0}^{\infty }dt^{\prime }\Phi _{\xi }\left( t^{\prime }\right).
\label{phase5}
\end{equation}

Notice that in the Continuous Time Random Walk (CTRW) as used in \cite
{klafter} yields, in the case where the waiting time distribution is
exponential, $\psi \left( t\right) =a\exp \left[ -at\right] $, the same
evolution for the probability density $p\left( x,t\right) $ as that for the
phase space distribution $\sigma \left( x,t\right) $ resulting from (\ref
{phase1}). This can be established by noticing that in the case of the
dichotomous variable $\xi $ used here, the waiting time distribution is
related to the correlation function by the exact relation \cite{geisel}

\begin{equation}
\Phi _{\xi }\left( t\right) =\frac{1}{\tau _{W}}\int_{t}^{\infty }dt^{\prime
}\left( t^{\prime }-t\right) \psi \left( t^{\prime }\right) ,  \label{phase4}
\end{equation}
where $\tau _{W}$ denotes the mean sojourn time. In the exponential case
this sojourn time becomes identical to the correlation time defined by Eq. (%
\ref{phase5}). In other words, since $\Phi _{\xi }\left( 0\right) =1$, in
the exponential case $a= \frac{1}{\tau _{W}}$ and $\tau _{C}=\tau _{W}$.

In this paper we study the case where the waiting time distribution $\psi(\tau)$ has the inverse power law form: 
\begin{equation}
\psi(\tau) = \frac{(\mu -1) T^{\mu -1}}{(t + T)^{\mu}},
\label{waitingtimedistribution}
\end{equation}
where $\mu=\beta +2$ and $T$ is a parameter determining the length of the laminar  region\cite
{allegrini}.  According to Eq. (\ref{phase4}), the corresponding correlation
function $\Phi_{\xi}(t)$  reads 
\begin{equation}
\Phi_{\xi}(t) = (\frac{T}{(T+t)})^{\beta}.  \label{correlationfunction}
\end{equation}
We shall focus our attention on $\beta < 1$. This means that the 
correlation time $\tau_{C}$ becomes infinite, while the mean waiting  time $%
\tau_{W} = T/\beta$ remains finite. As pointed out by the  authors of Ref.
\cite{mario}, the numerical calculation, based on  trajectories, yields a
perfect agreement with the predictions of  the GCLT at times 
$t >> \tau_{W}$.  This means that in this time scale the trajectory treatment  yields a
distribution that does not keep memory of the initial  condition.  This is
in a striking conflict with the non-Markovian character of  Eq. (\ref{phase1}%
), and already affords good reasons to explain  the emergence of two
distinct solutions from the same equation.  Actually, as we shall see, one
of these solutions is correct, and  it is the unique solution of Eq. (\ref
{phase1}). The other  solution rests on an approximation, necessary to
recover the  Markov nature of the process described by the GCLT. This is not
a genuine solution of Eq.  (\ref{phase1}): its departure from the exact
solution is a  measure of the discrepancy between trajectory and density.

It is worth pointing out that the CTRW of Ref.\cite{klafter} rests on the
waiting time distribution $\psi(t)$, whereas the generalized diffusion
equation of Eq. (\ref{lou6}) is based on the correlation function $%
\Phi_{\xi}(t)$. This explains why the former theory is compatible with the
Markov character of L\'{e}vy diffusion ($\tau_{W}$ is finite) while the
latter one is not ($\tau_{C}$ is infinite).

\section{On the correct solution to the generalized diffusion equation}

We shall investigate the solution to (\ref{phase1}) in the case where the
correlation time diverges. In the case where there is no correlation time,
the determination of the statistical properties of the system depends on
whether we analyze the trajectories or the density distribution function.
The two ''representations'' appear to give different results.

\subsection{Approximate solution}

The closed form solution to (\ref{phase1}) depends on the choice of the
correlation function in the integrand. We select an inverse power-law
correlation function for the kernel,

\begin{equation}
\Phi \left( t\right) =W^{2}\frac{T^{\beta }}{\left( T+t\right) ^{\beta }},
\label{sol1}
\end{equation}
with $0<\beta <1$, and $T$ is a positive constant. The equation for the
density distribution can now be written, with a little adjustment, as

\begin{equation}
\frac{\partial \sigma \left( x,t\right) }{\partial t}=W^{2}\frac{T^{\beta }%
} {\left( T+t\right) ^{\beta }}\int_{0}^{t}\frac{dt^{\prime }} {\left( 1-%
\frac{t^{\prime }}{T+t}\right) ^{\beta }}\frac{\partial ^{2}\sigma \left(
x,t^{\prime }\right) }{\partial x^{2}} ,  \label{sol2}
\end{equation}
where we have anticipated that since $t>t^{\prime }$, we can expand the
kernel using the binomial theorem. Retaining the lowest order term in $T$ \
in the binomial expansion we have

\begin{equation}
\frac{\partial \sigma \left( x,t\right) }{\partial t}\approx W^{2} \frac{%
T^{\beta }}{\left( T+t\right) ^{\beta }}\int_{0}^{t}\frac{\partial
^{2}\sigma \left( x,t^{\prime }\right) }{\partial x^{2}}dt^{\prime } ,
\label{sol3}
\end{equation}
so that defining

\begin{equation}
v\left( t\right) ^{2}\equiv W^{2}\frac{T^{\beta }}{\left( T+t\right) ^{\beta
}} ,  \label{sol4}
\end{equation}
we can transform the differential-integral equation (\ref{sol3}) into an
ordinary partial-differential equation by differentiating this equation in
time to obtain

\begin{equation}
\frac{\partial }{\partial t}\left[ \frac{1}{v\left( t\right) ^{2}}\frac{%
\partial \sigma \left( x,t\right) }{\partial t}\right] =\frac{\partial
^{2}\sigma \left( x,t\right) }{\partial x^{2}}.  \label{sol5}
\end{equation}
Introducing the dimensionless quantities $\tau =1+t/T$ and $q=x/WT,$ into (%
\ref{sol5}) and taking the Fourier transform with respect to the variable $q$
we arrive at Lommel's equation \cite{nikiforov}:

\begin{equation}
\frac{\partial ^{2}\widehat{\sigma }\left( k,\tau \right) }{\partial \tau
^{2}}+\frac{\beta }{\tau }\frac{\partial \widehat{\sigma }\left( k,\tau
\right) }{\partial \tau }+\frac{k^{2}}{\tau ^{\beta }}\widehat{\sigma }%
\left( k,\tau \right) =0 ,  \label{sol6}
\end{equation}
whose solution is expressed in terms of Bessel functions as:

\begin{equation}
\widehat{\sigma }\left( k,\tau \right) =\tau ^{\frac{1-\beta }{2}}\left[
a\left( k\right) J_{\nu }\left( \frac{2}{2-\beta }k\tau ^{\frac{2-\beta }{2}%
}\right) +b\left( \tau \right) J_{-\nu }\left( \frac{2}{2-\beta }k\tau ^{%
\frac{2-\beta }{2}}\right) \right] ,  \label{sol7}
\end{equation}
where the coefficients $a\left( k\right) $ and $b\left( \tau \right) $
satisfy the initial conditions:

\begin{equation}
\left. \widehat{\sigma }\left( k,\tau \right) \right| _{\tau =1}=1\,
{\rm and}\, \left. \frac{\partial \widehat{\sigma }\left( k,\tau \right) }{%
\partial \tau }\right| _{\tau =1}=0  \label{sol8}
\end{equation}
and $\nu =\frac{1-\beta }{2-\beta }.$ After some algebra we obtain

\begin{equation}
a\left( k\right) =-\frac{\frac{1-\beta }{2k}J_{-\nu }\left( \frac{2}{2-\beta 
}k\right) +J_{-\nu }^{^{\prime }}\left( \frac{2}{2-\beta }k\right) }{\sin
\nu \pi }\frac{\pi k}{2-\beta }  \label{sol9}
\end{equation}
and 
\begin{equation}
b\left( k\right) =\frac{\frac{1-\beta }{2k}J_{\nu }\left( \frac{2}{2-\beta }%
k\right) +J_{\nu }^{^{\prime }}\left( \frac{2}{2-\beta }k\right) }{\sin \nu
\pi }\frac{\pi k}{2-\beta }.  \label{sol10}
\end{equation}
Since the argument of the Bessel functions are almost always greater than $%
\nu $ we can adopt the approximation

\begin{equation}
J_{\nu }\left( z\right) \approx\sqrt{\frac{2}{\pi z}}\sum_{n=0}^{m-1}\frac{%
c\left( n\right) }{z^{n}}\cos \left[ z-\frac{\pi \left( 2\nu -2n+1\right) }{4%
}\right],  \label{sol11}
\end{equation}
which when inserted into the solution (\ref{sol7}) enables us to invert the
Fourier transform.

The density distribution resulting from the Fourier inversion of this lowest
order solution to the Liouville equation, remembering that $q$ is the
Fourier complement to $k$, is

\begin{equation}
\sigma \left( x,t\right) =\frac{T^{\beta /4}}{2\left( T+t\right) ^{\beta /4}}%
\delta \left[ \frac{\left| x\right| }{WT}-\frac{2}{2-\beta }\left( \frac{%
T^{1-\beta /2}}{\left( T+t\right) ^{1-\beta /2}}-1\right) \right] ,
\label{sol12}
\end{equation}
with the region between the two peaks corresponding to a negligible constant
density distribution. What is interesting about this solution is that it has
two delta function peaks traveling in opposite directions at a speed which
is time dependent. It is a simple matter to check that for early times, $T>>t
$, we have a ballistic propagation front that is characteristic of a system
of trajectories with $x\sim \pm Wt.$ On the other hand, for late times, $t>>T
$, the distribution has propagation fronts that scale with time as $x\sim
\pm t^{1-\beta /2}$, whereas the trajectories themselves would still
maintain the ballistic front.

So what have we learned? In the exponential case the trajectory and the
density pictures coincide. Now we have also established that for early
times, $T>>t,$ there is no difference in the evolutions of the system
between the exponential and inverse power-law correlation functions. This is
the reason why the two pictures coincide. However, at late times, when the
memory becomes important, the scaling of the trajectory and density pictures
appear to be quite different. But this was an approximate solution, what
about the exact solution?

\subsection{The exact solution}

In this section we are interested in the asymptotic behavior of the exact
solution to (\ref{phase1}). The most direct way to determine these
properties is to take the Laplace transform in time and Fourier transform in
space to obtain the Fourier-Laplace transform of the Liouville density

\begin{equation}
\widehat{\widetilde{\sigma }}\left( k,s\right) =\frac{1}{s+\widetilde{\Phi }
\left( s\right) k^{2}} ,  \label{sol13}
\end{equation}
where we have imposed the initial conditions

\begin{equation}
\left. \sigma \left( x,t\right) \right| _{t=0}=\delta \left(
x\right)  
{\rm and} \left. \frac{\partial \sigma \left( x,t\right) }{
\partial t}\right| _{t=0}=0.  \label{sol14}
\end{equation}
The inverse Fourier transform of (\ref{sol13}) yields

\begin{equation}
\widetilde{\sigma }\left( x,s\right) =\sqrt{\frac{s}{\widetilde{\Phi }\left(
s\right) }}\frac{e^{-\left| x\right| \sqrt{\frac{s}{\widetilde{\Phi }\left(
s\right) }}}}{2s} ,  \label{sol15}
\end{equation}
which we can integrate to obtain

\begin{equation}
\int_{-\infty }^{\infty }\widetilde{\sigma }\left( x,s\right) dx=\frac{1}{s},
\label{sol16}
\end{equation}
indicating the conservation of normalization over time. To go beyond the
formal solution (\ref{sol15}) we must specify the correlation function. We
assume the inverse power-law form given by (\ref{sol1}) so that its Laplace
transform becomes

\begin{equation}
\widetilde{\Phi }\left( s\right) =\frac{\Gamma \left( 1-\beta \right) TW^{2}%
}{\left( sT\right) ^{1-\beta }}\left[ e^{sT}-E_{\beta -1}^{sT}\right],
\label{sol17}
\end{equation}
where the generalized exponential function is defined by \cite{bologna,west}

\begin{equation}
E_{\gamma }^{x}\equiv D_{x}^{\gamma }\left[ e^{x}\right] =\sum_{n=0}^{\infty
}\frac{x^{n-\gamma }}{\Gamma \left( n+1-\gamma \right) } .  \label{sol18}
\end{equation}

\subsubsection{Early time behavior}

Let us first consider the behavior of the correlation function at early
times. In this domain, $t\rightarrow 0$, we have $s\rightarrow \infty $, so
that the generalized exponential becomes

\begin{equation}
E_{\beta -1}^{sT}\approx e^{sT}-\frac{1}{\Gamma \left( 1-\beta \right)
\left( sT\right) ^{\beta }},  \label{sol19}
\end{equation}
which when substituted into (\ref{sol17}) yields $\widetilde{\Phi }\left(
s\right) \approx W^{2}/s$ so that the Laplace transform of the early time
approximation to the exact solution is

\begin{equation}
\widetilde{\sigma }\left( x,s\right) \approx \frac{e^{-\left| x\right| s/W}}{%
2W}.  \label{sol20}
\end{equation}
The inverse Laplace transform of (\ref{sol20}) yields the delta function

\begin{equation}
\sigma \left( x,t\right) \approx \frac{1}{2W}\delta \left( t-\frac{\left|
x\right| }{W}\right) .  \label{sol21}
\end{equation}
Thus, for times shorter than $T$, the evolution of the Liouville density
consists of two peaks traveling in opposite directions at the same speed, $%
W $. Note that this is the same early-time solution obtained in the previous
section.

\subsubsection{Late time behavior}

Now let us consider the asymptotic in time behavior of the exact solution.
In the late time domain $t\rightarrow \infty ,$ we have $s\rightarrow 0$, so
examining the behavior of (\ref{sol17}) in this domain we have

\begin{equation}
\widetilde{\Phi }\left( s\right) \approx \frac{\Gamma \left( 1-\beta \right)
TW^{2}}{\left( sT\right) ^{1-\beta }}\left[ 1-\frac{\left( sT\right)
^{1-\beta }}{\Gamma \left( 1-\beta \right) }\right] .  \label{sol22}
\end{equation}
Note that as $s\rightarrow 0$ the leading term in this expansion diverges
for $\beta <1$ corresponding to the fact that there is no correlation time
for this process. Inserting this expression for the Laplace transform of the
correlation function into (\ref{sol17}), keeping only the diverging term,
yields

\begin{equation}
\widetilde{\sigma }\left( x,s\right) \approx \frac{\exp \left[ -\frac{\left|
x\right| s^{1-\beta /2}}{\Gamma \left( 1-\beta \right) WT^{\beta /2}}\right] 
}{2\Gamma \left( 1-\beta \right) W\left( sT\right) ^{\beta /2}}.
\label{sol23}
\end{equation}
The inverse Laplace transform of (\ref{sol23}) is

\begin{equation}
\sigma \left( x,t\right) \approx \frac{1}{2\left\langle x\right\rangle
t^{1-\beta /2}}\sum_{n=0}^{\infty }\frac{\left( -1\right) ^{n}}{n!\Gamma
\left( 1-\left( n+1\right) \left( 1-\beta /2\right) \right) }\left( \frac{%
\left| x\right| }{\left\langle x\right\rangle t^{1-\beta /2}}\right) ^{n} ,
\label{sol24}
\end{equation}
where the average of the system variable is

\begin{equation}
\left\langle x\right\rangle =WT^{\beta /2}\Gamma \left( 1-\beta \right),
\label{sol25}
\end{equation}
as had been obtained previously \cite{barkai}. Straight forward dimensional
analysis indicates that the space variable scales as $x\sim t^{\alpha }$,
where

\begin{equation}
\alpha =1-\beta /2 . \label{sol26}
\end{equation}
This
scaling is the same scaling as that of the second moment\cite
{allegrini,trefen} and it is different from the L\'{e}vy scaling, which is, as we shall see in Section IV,
$\alpha = 1/(\beta +1)$. Thus, in the late time region the exact solution of Eq. (\ref{lou6}) departs from 
the L\'{e}vy diffusion, in accordance with the conclusions of the authors of Ref. \cite{metzler}.
It is worth remarking that from a mathematical point of view, according to
Ref.\cite{weiss}, the equation of motion admits a solution also for $|x| >>
t^{\alpha}$.This solution is proved to drop exponentially with $|x|$. We
think that the two peaks traveling in the opposite direction of Eq.(\ref
{sol12}) are a signature of the two ballistic peaks stemming from the
numerical calculation based on trajectory dynamics\cite{mario,allegrini}.
The population beyond these two peaks is rigorously zero. On the basis of
this physical suggestion we neglect this contribution to the general solution
of Eq. (\ref{lou6}). This makes the solution proposed in this paper different from that 
proposed by the authors of Ref. \cite{metzler}. However, this is not relevant for the main finding of this paper, the conflict between trajectory and density perspective, since
also the solution of the authors of Ref. \cite{metzler} departs from L\'{e}vy statistics. We only note that our choice insures the existence of a unique scaling while
the solution of the authors of Ref. \cite{metzler} does not. 
To support our conclusion we note that in the asymptotic limit $s \rightarrow 0, k \rightarrow 0$, Eq. (\ref{sol13}) yields
\begin{equation}
\label{heuristicscaling}
 {\hat {\tilde \sigma}}(k,s) = \frac{1}{s + {\rm const} \, s^{\beta -1} k^{2}}.
\end{equation}
The scaling condition $x \sim t^{\alpha}$ implies $k = s^{\alpha}$, which plugged into the right hand side term of Eq.(\ref{heuristicscaling}) makes the left hand side of the same equation proportional to $1/s$ when the scaling condition of Eq. (\ref{sol26}) applies. We note that $1/s$ is the Laplace transform of a constant in accordance with the fact that scaling is a reflection of stationarity. For this reason we are 
inclined to believe that the density perspective yields in the asymptotic limit a unique scaling and that our solution correctly reflects this condition.

\section{L\'{e}vy scaling}

It is clear that the memory effect contained in the correlation function,
that is, the memory kernel in (\ref{phase1}), implies that the density
dynamics are not Markovian. However, as earlier pointed out, the
individual trajectories
are driven by the waiting time distribution $\psi(\tau)$, and
consequently, in accordance with the GCLT,
yield the Markov condition under the form of L\'{e}vy statistics.
Here we discuss how to make Eq. (\ref{phase1})
compatible with this Markov condition. This requires the adoption of
an unusual form of a Markov approximation. First of all we change the notation from $\sigma$ to $p$ to stress that we depart from 
a rigorous treatment in terms of the density $\sigma$. Then we identify $p(x,t-t')$ with the change to the probability
distribution $p(x)$ occurring when time changes from $t^{\prime}$ to $t$, $\Delta p\left( x,t;t^{\prime }\right)$.We
assume this change to fit the following relation

\begin{displaymath}
\Delta p\left( x,t;t^{\prime }\right) =\frac{1}{2W}\int_{-\infty }^{\infty
}\delta \left[ t^{\prime }-\frac{\left| x-x^{\prime }\right| }{W}\right]
p\left( x^{\prime },t\right) dx^{\prime }
\end{displaymath}

\begin{equation}
-\frac{1}{2W}\int_{-\infty }^{\infty }\delta \left[ t^{\prime }-
\frac{%
\left| x-x^{\prime }\right| }{W}\right] p\left( x,t\right) dx^{\prime }.
\label{levy1}
\end{equation}
The total change is zero, as it can be easily 
assessed by integrating $\Delta p\left( x,t;t^{\prime }\right)$ from $-\infty$ to $+\infty$. This is so because the change is assumed to be determined only  by the uniform motion with velocity $W$. 
We impose the condition (\ref{levy1}) on the form of the Liouville densities in (\ref{phase1}). By
integrating out the delta functions, and doing some algebra, we obtain the
master equation \cite{bologna}

\begin{displaymath}
\frac{\partial p\left( x,t\right) }{\partial t}=\frac{1}{2W}\int_{-\infty}^{\infty}\left[ \frac{%
\partial ^{2}}{\partial x^{\prime 2}}\Phi \left( \left| x-x^{\prime }\right|
\right) \right] p\left( x^{\prime },t\right) dx^{\prime }
\end{displaymath}
\begin{equation}
-\frac{1}{2W}\int_{-\infty}^{\infty}\left[
\frac{\partial ^{2}}{\partial x^{\prime 2}}\Phi \left( \left| x-x^{\prime
}\right| \right) \right] p\left( x,t\right) dx^{\prime }.  \label{levy2}
\end{equation}
Inserting the inverse power-law correlation function into (\ref{levy2}),
performing the indicated differentiations and taking the limit $kTW<<1$ has
been shown to yield \cite{bologna}
\begin{equation}\frac{\partial p\left( x,t\right) }{\partial t}=b\int_{-\infty }^{\infty }%
\frac{p\left( x^{\prime },t\right) dx^{\prime }}{\left| x-x^{\prime }\right|
^{\beta +2}} ,
 \label{levy3}
\end{equation}
which is the Reisz fractional derivative discovered by Seshadri and West 
\cite{west2} and whose solution is the symmetric L\'{e}vy distribution. The
approximation of Eq. (\ref{levy1}) makes it possible for us to get rid of
the time convolution nature of the generalized diffusion equation of Eq.(\ref
{lou6}). At the same time, this key relation replaces the correlation
function $\Phi_{\xi}(t)$ with its second-order derivative, and,
consequently, in accordance with Eq. (\ref{phase4}), with the waiting time
distribution $\psi(\tau)$.

The symmetric L\'{e}vy distribution that solves (\ref{levy3}) is

\begin{equation}
p\left( x,t\right) =\int_{-\infty }^{\infty }e^{ikx}e^{-bt\left| k\right|
^{\beta +1}}\frac{dk}{2\pi } ,  \label{levy4}
\end{equation}
which satisfies the scaling relation

\begin{equation}
p\left( x,t\right) =\gamma ^{\frac{1}{\beta +1}}p\left( \gamma ^{\frac{1}{%
\beta +1}}x,\gamma t\right) .  \label{levy5}
\end{equation}
From (\ref{levy5}) it is clear that the L\'{e}vy diffusion process has a
scaling, $x\sim t^{\alpha }$, with

\begin{equation}
\alpha =\frac{1}{\beta +1}.  \label{levy6}
\end{equation}
This scaling is consistent with the process generated by the fluctuations of
the variable $\xi $, as proved by the numerical simulation \cite{mario}. It
is evident that at a given time $t$, such that $t>>t_{W}$, a single
trajectory has been sojourning in a given laminar phase a number of times
given by

\begin{equation}
N=\frac{t}{t_{W}}.  \label{levy7}
\end{equation}
It is also evident that the position occupied by the diffusing particle at
time $t$ is equivalent to the superposition of $N$ variables $y_{j}$, $%
j=1,2,...,N$, each with the same inverse power-law waiting time distribution
and index $\beta +2.$ In fact, the extended time regions, of duration $%
\tau_{j}$, where the variable $\xi$ keeps a constant value, either $W$ or $-W
$, corresponds to uncorrelated ``flights'' of length $|y_{j}| = W\tau_{j}$.
Thus, the GCLT applies to the superposition of trajectories, resulting in a
L\'{e}vy process, and we have the scaling with the power-law index given by (%
\ref{levy5}). However, this solution, which agrees with the results of
numerical simulation and the predictions of the GCLT, and so with the
trajectory perspective, dramatically departs from the unique solution of Eq.
(\ref{lou6}), illustrated in Section III.

What is the reason for this conflict? The numerical treatment\cite
{allegrini,mario} shows that there exists a multi-scaling condition. In
fact, there is a finite probability that a trajectory, with velocity $W$, at 
$t= 0$, maintains this velocity up to a given time $t$. The number of these
trajectories is proportional to the correlation function $\Phi_{\xi}(t)$,
and consequently decays as $1/t^{\beta}$, with an infinite decay time. These
trajectories generates the two side peaks of the broadening distribution,
and a ballistic propagation front. This means that the probability
distribution does not have a defined scaling, but it is rather the
combination of two. The probability distribution, enclosed by the two
ballistic peaks is proven numerically\cite{mario,allegrini} to have the
scaling $\alpha = 1/(\beta + 1)$, namely, the L\'{e}vy scaling of Eq.(\ref
{levy6}). The propagation front, moving ballistically, would yield the
scaling $\alpha = 1$. The exact solution of Section III, as we have seen,
yields the scaling of Eq. (\ref{sol26}), which is a kind of compromise
between the ballistic and the L\'{e}vy scaling, being smaller than the
former and greater than the latter. It seems that the Liouville-like
approach is unable to reproduce this multiple scaling condition, and this is
probably the reason of conflict between trajectory and density picture.

\section{First Consequence: foundation of statistical mechanics}

The foundation of statistical mechanics is still the object of debates, and
it seems to us that the main conflict is that between the advocates of
trajectory randomness and the advocates of the infinitely large number of
degrees of freedom, as main ingredient to produce the transition from
dynamics to thermodynamics. As an example of the former view we quote Ref.
\cite{bianucci}. In this paper, although the importance of using a very
large number of degrees of freedom for the foundation of ordinary
statistical mechanics is not ruled out, deterministic chaos of classical
trajectories is suggested to be the main ingredient for the foundation of
statistical mechanics. The authors of this paper\cite{bianucci} derive
indeed for temperature an expression more general than that proposed by
Boltzmann, and recover the Boltzmann principle in the limiting case of
infinite degrees of freedom.

The advocates of infinitely many degrees of freedom, as main ingredient for
the foundation of statistical mechanics, are many. The first, as repeatedly
stated by Lebowitz, is Boltzmann. In the last few years the traditional
point of view of Boltzmann has been sustained by Lebowitz\cite{lebowitz} and
by Goldstein \cite{goldstein}. Another aspect of the controversy has to do
with the origin itself of entropy production. The former party seems to rest
on the impossibility of keeping under control the initial conditions not so
much as a consequence of trajectory instability, but much more as a
consequence of the infinitely many degrees of the systems that are expected
to characterize the systems that at the macroscopic level exhibit
thermodynamic properties. In a sense, the conflict between these two
distinct perspectives might lead the advocates of the latter party\cite
{lebowitz,goldstein} to support the derivation of the generalized master
equation here under study through the procedure adopted in Section II, with
the relaxation process being induced by the action of infinitely many
degrees of freedom. The advocates of randomness would probably make the
choice of the derivation of Appendix. The two views yield equivalent results
in the case of ordinary statistical mechanics. In fact the Fick's law
derived in Ref. \cite{lee} without involving trajectory randomness is
compatible with the CLT accounting for the Gaussian nature of the ordinary
diffusion processes. In the case of strange kinetics here under discussion
the advocates of randomness, as a source of memory erasure, would probably
adopt the Markov approximation of Section IV. This approximation has in fact
the attractive property of establishing a perfect agreement with numerical
simulation. However, in so doing, the advocates of randomness should
acknowledge that the adoption of the generalized Master equation, here under
discussion, is not appropriate and that L\'{e}vy processes are incompatible
with a Liouville-like approach.

The experimental assessment of L\'{e}vy statistics might be regarded as a
compelling evidence of the existence of random trajectories. These random
trajectories, on the other hand, are characterized by a well defined
Kolmogorov-Sinai (KS) entropy\cite{rosa}. In fact, the diffusing
trajectories are generated by an intermittent process, and the entropy
increase is generated by the random choice of the waiting times of Eq. (\ref
{waitingtimedistribution}). Since the dynamic realization of L\'{e}vy
diffusion, with $2 < \mu < 3$ implies the existence of a finite $\tau_{W}$,
the entropy increase per unit of time is fixed and a departure from this
ordinary KS condition is expected to take place only for $\mu < 2$, a
condition that would yield $\tau_{W} = \infty$.

\section{Second Consequence: Decoherence and spontaneous wave-function
collapses}

In a recent paper, celebrating 100 years of the quantum\cite{arcibald},
Tigmark and Wheeler emphasize that the de-coherence theory has made obsolete
the hypothesis of the wave-function collapses of the founding fathers of
quantum mechanics. A classical trajectory is imagined as a succession of
dimensionless points, with the particle being located at given time in only
one of these positions. This is quite different from a wave function that
might imply the presence of a particle at the same time in different
positions, say, two distinct positions. According to de-coherence theory\cite
{zeh} the classical trajectories are recovered because the environment
measures, so to speak, the position, and provokes a collapse into one of the
two positions. As Tigmark and Wheeler pointed out, this is not a real
collapse but only a consequence of the system entanglement with the
environment. The reduced density matrix "looks like" the one that would be
produced by a real collapse, and while we have the impression of seeing a
collapse, actually there has been an entanglement between system and
environment, driven by a unitary transformation. This simulacrum of a
collapse apparently conflicts with a realistic perspective, but d'Espagnat
in a recent paper\cite{bernard} warns us that the answer to the question
"How do we know that there is a stone on the path, or a tree in the
courtyard?" must be addressed with caution. He urges us to adopt this
approach: "We know that if had a look at the path, to check whether or not
we have the impression of seeing a stone, we should actually get the
impression in question." It is difficult to support the realistic
perspective advocated by Bassi and Ghirardi\cite{bassi} precisely because
wave-function collapses and de-coherence theory produce the same reduced
density matrix.

The result of this paper might transform  the philosophical debate between 
d'Espagnat and Bassi and Ghirardi into a real  scientific issue. Our
arguments are conjectural but  plausible. Let us see why. Let us consider
Eq. (\ref{phase1}). If we move from within a quantum mechanical picture, we
have to make the conjecture that the derivation of the classical process of
anomalous diffusion must go through this equation. In other words, if the
de-coherence program is reliable, and yields any kind of classical
diffusion, normal and anomalous, then we expect that it will do it remaining
at the level of densities, without ever invoking the concept of trajectory.
Of course, the Markovian approximation stemming from Eq.(\ref{levy1}) would
not be possible and we should therefore give credit to the exact solution of
Eq. (\ref{phase1}). If, on the contrary, the trajectories exist and are
produced by real wave-function collapses, then we expect that the L\'{e}vy
statistical mechanics might emerge. We would be tempted to say that the
experimental assessment of the existence of L\'{e}vy processes is a proof of
the occurrence of genuine wave-function collapses. It is evident, however,
that this would require a more careful derivation from Hamiltonian dynamics,
and the experimental assessment of L\'{e}vy processes with a microscopic
rather than a macroscopic and phenomenological origin.

\section{Concluding Remarks}

Sections V and VI refer to possible consequences of the results of this
paper that are very ambitious. In the final balance of this Section we adopt
a much more modest view: we make a short history of the facts that led us to
reach the conclusions illustrated in this paper. A nice dynamic foundation
of L\'{e}vy processes was given by Zumofen and Klafter\cite{klafter}. The
source of fluctuations used by these authors is very similar to that adopted
by us in Appendix. The theory used by them to account for their experimental
results is that of CTRW. As already remarked in Section III, this theory is
not based on the Liouville-like prescriptions used in this paper, and, it
is, in our view compatible with the concept of trajectory. This is why the
authors of Ref. \cite{klafter} found correct and non-contradictory results.
The authors of Ref. \cite{allegrini}, on the contrary, adopted a
Liouville-like approach that led them to establish, for the first time, the
generalized diffusion equation here under discussion. However, these authors
made use of computer simulation and found that the Markov approximation
of Section IV is necessary to establish a satisfactory accordance with the
results of numerical simulation. They made also the wrong conjecture that
the Markov approximation of Eq. (\ref{levy1}) is legitimate from a
mathematical as well as from a physical point of view, and that the solution
emerging from it is very close to the exact solution, unknown to them. This
is the main reason why the later paper of Ref.\cite{obscure} was fraught by
internal contradictions, correctly pointed out by Metzler and Nonnemacher
\cite{metzler} in a subsequent paper. The authors of Ref. \cite{bologna},
having in mind the numerical results of the earlier work\cite{allegrini},
used the generalized master equation\cite{kenkre}, as a bridge between the
Hamiltonian microscopic dynamics and the long-time limit, or macroscopic
level, where L\'{e}vy statistics show up. The physical arguments are correct
and the agreement with the numerical results is remarkable. However, no
attention was devoted by these authors\cite{bologna} to the striking fact
that the generalized diffusion equation (\ref{lou6}) is incompatible with
the Markovian approximation necessary to derive L\'{e}vy statistics. We hope
that the present paper might serve at least the good purpose of explaining
the mathematical and physical reasons behind the contradictory conclusions
of the papers of Refs.\cite{allegrini,metzler,obscure,bologna}. We think
that this might bear also the significant consequences of establishing the
borders between dynamics and thermodynamics (Section IV) and between quantum
and classical mechanics (Section V). This will be the object of further
research work.

\setcounter {equation} {0} 
\renewcommand{\theequation}{A-\arabic{equation}} 
\section{ APPENDIX: Frobenius-Perron operator for an idealized model of
intermittent dynamics} $ 
$ Let us consider the following dynamical system. This is given by a
coordinate $y$ moving within the interval $I = [0,2]$ with the equation of
motion 
\begin{equation}
\frac{dy}{dt} = -\frac{dV(y)}{dy} = -W(y).  \label{intermittentmotion}
\end{equation}
This is a form of overdamped dynamics within the "potential" $V(y)$, with
the function $W(y)$ defined by 
\begin{equation}
W(y) = {\rm sign}(y-1)\lambda |y-1|^{z},  \label{left2}
\end{equation}
where ${\rm sign(x)}$ denotes the sign of $x$. 
The particle with coordinate $y$ moves within a potential with the minimum
located at $y= 1$. Thus, if the initial condition of the particle is $y(0) >
1$, the particle moves from the right to the left towards the potential
minimum. If the initial condition is $y(0) < 1$, then the motion of the
particle towards the potential minimum takes place from the left to the
right. When the particle reaches the potential bottom is injected to an
initial condition, different from $y=1$, chosen in a random manner. We thus
realize a mixture of randomness and slow deterministic dynamics. The left
and the right portions of the potential $V(y)$ correspond to the laminar
regions of turbulent dynamics, while randomness is concentrated at $y = 1$.
It is straightforward to prove that the waiting time distribution in any of
the two laminar regions is given by

\begin{equation}
\psi(\tau) = \frac{(\mu-1) T^{\mu -1}} {(T+\tau)^{\mu}},
\label{waitingtimes}
\end{equation}
where $\mu = z/(z-1)$ and $T = (\mu-1)/\lambda$. Eq.(\ref{lou1}) must be
written now as 
\begin{equation}
\frac{dx}{dt} = \xi(y(t)),  \label{newform}
\end{equation}
with $\xi(y) = W$ if $1< y \leq 2$,and $\xi(y) = -W$ if $0< y \leq 1$. The
hamiltonian formulation corresponding to this kind of sporadic randomness
with an inverse power law distribution has been discussed in detail by
Zaslavsky\cite{zaslavsky}. Unfortunately, this Hamiltonian treatment would
make the calculations more complicated. We think that the current treatment,
although not Hamiltonian, shares the main dynamic properties of Zaslavsky's
dynamics. In this idealization the size of the chaotic region is reduced to $%
y=1$, with a virtually instantaneous back injection into the laminar region.
In the Hamiltonian models of Zaslavsky, on the contrary, the particle spends
some time in the chaotic region, with no significant consequences on the
asymptotic time properties, though, since these are dominated by the laminar
motion, due to the inverse power law nature of the waiting time distribution.

In this case the Liouville-like equation for the whole phase space $(x,y)$
reads 
\begin{equation}  \label{total}
\frac{\partial}{\partial t} \phi(x,y,t) = \hat{L}_{T}\phi(x,y,t),
\end{equation}
where the Frobenius-Perron operator of the whole system, $\hat{L}_{T} $, is
given by 
\begin{equation}
\hat{L}_{T} \equiv - \xi(y) \frac{\partial}{\partial x} + \hat{L}_{B} .
\end{equation}
The first term on the right hand side of this equation represents the
interaction between the diffusing variable $x$ and the fluctuation $\xi$
driven by the intermittent dynamics of the variable $y$. The
Frobenius-Perron operator corresponding to this "bath" dynamics is given by 
\begin{equation}  \label{bathdynamics}
\hat{L}_{B} [.] \equiv \frac{\partial}{\partial y} V(y) [.] + \frac{1}{%
\tau_{random}} \int_{0}^{2} dy \delta (y -1) [.] .
\end{equation}
We assign to the back injection a finite time $\tau_{random}$, whose actual
value is not important for the present discussion, provided that it is
assumed to be very short compared to the mean waiting time $\tau_{W}$. The
second term on the right hand side of Eq. (\ref{bathdynamics}) becomes
active only if the probability distribution does not vanish for $y = 1$. In
that case it contributes an increase of the population at any value $y \neq 1
$, with uniform probability. If it vanishes, the time evolution of the
distribution density is not affected by the back injection. When the
condition of equilibrium are reached, and the distribution of $y$ is at
equilibrium the process of back injection becomes time independent\cite{rosa}%
. Using the method of analysis developed in Ref. \cite{rosa}, it is possible
to find the bath equilibrium distribution $\eta(y)$, namely, the
distribution fulfilling the condition 
\begin{equation}  \label{equilibriumofb}
\hat{L}_{B}\eta(y) = 0 .
\end{equation}
We are now equipped to adopt the Zwanzig projection method\cite{grigo}. The
projection operator $P$ works according to the following prescription 
\begin{equation}
P\phi(x,y,t) = \sigma(x,t) \eta(y).  \label{zwanzigprojection}
\end{equation}
Let us assume the factorized initial condition: $\phi(x,0) = \sigma(x,0) \eta(y)$. 
The Zwanzig projection method yields the following reduced equation of motion

\begin{equation}
\frac{\partial}{\partial t}\sigma(x,t) = \frac{1}{\eta(y)} \int_{0}^{t} P 
\hat{L}_{T}e^{Q\hat{L}_{T}(t-t^{\prime})} Q \hat{L}_{T}
P\phi(x,y,t^{\prime})dt^{\prime} , \label{zwanzigprojectionmethod}
\end{equation}
where $Q \equiv 1-P$. 
By exploiting the dichotomous nature of the variable $\xi$, it is
straightforward to prove that Eq.(\ref{zwanzigprojectionmethod}) yields Eq.(%
\ref{lou6}). This confirms that this generalized diffusion equation is the
proper representation of the dynamic process under study. The conflict
between trajectory and density perspective cannot be ascribed to a
questionable use of the rules currently adopted to establish the connection
between density and trajectory perspective.



\begin{thebibliography}{}

\bibitem{petrosky}  T. Petrosky and I. Prigogine, Chaos, Solitons \&
Fractals {\bf 7}, 441 (1996).

\bibitem{petrosky2}  T. Petrosky and I. Prigogine, Chaos, Solitons \&
Fractals {\bf 11}, 373 (2000).

\bibitem{lee}  M.H. Lee, Phys. Rev. Lett. {\bf 85}, 2422 (2000).

\bibitem{kolmogorov}  Gnedenko and Kolmogorov, {\it Limit Distributions for
Sums of Independent Random Variables}, Addison-Wesley Pub., Cambridge (1954).

\bibitem{allegrini}  P. Allegrini, P. Grigolini and B.J. West, Phys. Rev. E 
{\bf 54}, 4760 (1996).

\bibitem{ronald}  R.F. Fox, J. Math. Phys. {\bf 18}, 2331 (1977).

\bibitem{simon}  B. Simon, arXiv:math-ph/9906008.

\bibitem{klafter}  G. Zumofen and J. Klafter, Phys Rev. E {\bf 47}, 851
(1993).

\bibitem{geisel}  T. Geisel, in {\it L\'{e}vy Flights and Related Topics in
Physics}, Proceedings, Nice, France ; Editors, M.F. Shlesinger, G.M.
Zaslavsky, U. Frisch, Lecture Notes in Physics {\bf 450}, 153 (1995).

\bibitem{mario}  M. Annunziato, P. Grigolini,  Phys. Letters A {\bf 269}, 31
(2000).

\bibitem{nikiforov}  A.F. Nikiforov and V.B. Uvarov, {\it Special Functions
of Mathematical Physics}, Transl. by R.P. Boas, Boston, Birkha\"{u}ser
(1988).

\bibitem{bologna}  M. Bologna, P. Grigolini and J. Riccardi, Phys. Rev. E 
{\bf 60}, 6435 (1999).

\bibitem{west}  B.J. West, M. Bologna and P. Grigolini, {\it Calculus of
Complexity} (unpublished)

\bibitem{barkai}  E. Barkai, R. Metzler and J. Klafter, Phys. Rev. E {\bf 61}%
, 132 (2000).

\bibitem{metzler}  R. Metzler and T.F. Nonnenmacher, Phys. Rev. E {\bf 57},
6409 (1998).

\bibitem{trefen}  G. Tref\'{a}n, E. Floriani, B.J. West and P. Grigolini,
Phys. Rev. E \ , 2564 (1994).

\bibitem{weiss}  G.H. Weiss, {\it Aspects and Applications of the Random Walk}, North-Holland, (1994).

\bibitem{west2}  V. Seshadri and B.J. West, Proc. Natl.  Acad. Sci. U.S.A. {\bf 79}, 4051 (1982).

\bibitem{bianucci}  M. Bianucci, R. Mannella, B.J. West, P. Grigolini, Phys.
Rev. E {\bf 51}, 3002 (1995).

\bibitem{lebowitz}  J.L. Lebowitz, Rev. Mod. Phys. {\bf 71} S346 (1999);
Physica A {\bf 263}, 516 (1999)

\bibitem{goldstein}  S. Goldstein, arXiv:cond-mat/0106496.

\bibitem{rosa}  M. Ignaccolo, P. Grigolini, and A. Rosa, Phys. Rev. E, {\bf %
64}, 026210 (2001).

\bibitem{arcibald}  M. Tigmark, J.A. Wheeler, Scientific American, Feb.
2001, p. 68, arXiv:quant-ph/0101077

\bibitem{zeh}  D. Giulini, E. Joos, C. Kiefer, J.Kupsch, I.-O. Stamatescu,
H.D. Zeh, {\em Decoherence and the Appearance of a Classical World in
Quantum Theory}, Springer, Berlin (1996).

\bibitem{bernard}  B. d'Espagnat, Phys. Lett. A, {\bf 282}, 133 (2001).

\bibitem{bassi}  A. Bassi and G.C. Ghirardi, Phys. Lett. A, {bf 282}, 133
(2001).

\bibitem{obscure}  B.J. West, P. Grigolini, R. Metzler, T. Nonnenmacher,
Phys. Rev. E {\bf 55}, 99 (1997).

\bibitem{kenkre}  V.M. Kenkre, E.W. Montroll, and N.F. Shlesinger, J. stat.
Phys. {\bf 9}, 3964 (1989).

\bibitem{zaslavsky}  G.M. Zaslavsky, {\em Physics of Chaos in Hamiltonian
systems}, Imperial College Press, London (1998).



\bibitem{grigo}  P. Grigolini, Adv. Chem. Phys. {\bf 62}, 1 (1985).
\end{thebibliography}
\end{document}